\documentclass[aps,prl,twocolumn,showpacs,amsmath,amssymb]{revtex4}
\usepackage{verbatim}
\usepackage[T1]{fontenc}
\usepackage{braket}
\usepackage{graphicx}
\usepackage{epstopdf}
\newcommand{\bdS}{\ensuremath{\boldsymbol{S}}}
\newcommand{\bdF}{\ensuremath{\boldsymbol{F}}}
\newcommand{\I}{\ensuremath{\mathrm{i}}}
\newcommand{\bdXi}{\ensuremath{\boldsymbol{\Xi}}}
\newcommand{\e}{\ensuremath{\mathrm{e}}}

\newcommand{\Ref}[1]{(\ref{#1})}
\newcommand{\bgeq}{\begin{equation}}
\newcommand{\eeq}{\end{equation}}

\newcommand{\lt}{\left }\newcommand{\rt}{\right}
\begin{document}

\title{From physical linear systems to discrete-time series. \\
A guide for analysis of the sampled experimental data}






%


\author{Jakub \'Sl\k{e}zak}
\email{jakub.slezak@pwr.edu.pl}
\author{ Aleksander Weron}
\email{aleksander.weron@pwr.edu.pl}
\affiliation{Hugo Steinhaus Center at Wroc{\l}aw University of Technology,\\ 50-370 Wroc{\l}aw, Poland}
\homepage{http://prac.im.pwr.wroc.pl/~hugo}

\date{\today}

\begin{abstract}
 Modelling physical data with linear discrete time series, namely Fractionally Integrated  Autoregressive Moving Average (ARFIMA), is a technique which achieved attention in recent years. However, these models are used mainly as a statistical tool only, with weak emphasis on physical background of the model. The main reason for this lack of attention is that ARFIMA model describes discrete-time measurements, whereas physical models are formulated using continuous-time parameter. In order to remove this discrepancy we show that time series of this type can be regarded as sampled trajectories of the coordinates governed by system of linear stochastic differential equations with constant coefficients. The observed correspondence provides formulas linking ARFIMA parameters and the coefficients of the underlying physical stochastic system, thus providing a bridge between continuous-time linear dynamical systems and ARFIMA models. 
\end{abstract}


\maketitle

\begin{figure}[h!] \centering
\includegraphics[width=8cm]{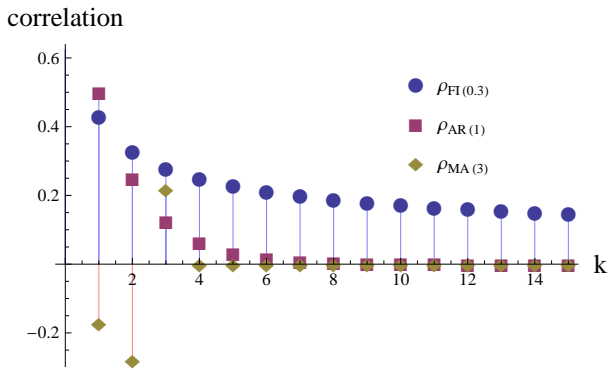}
\caption{(Color online) Memory functions of ARMA and ARFIMA processes: power-law (circles), exponential (squares) and finite (diamonds).}\label{f:corr}
\end{figure}

\begin{center}
\textbf{I. INTRODUCTION}
\end{center}
Discrete-time series methods based on ARMA (Autoregressive Moving Average) model, and more recently, its generalisation ARFIMA (Fractionally Integrated ARMA, also called FARIMA) \cite{granger,BDbook, beran} provide powerful and flexible statistical tools which were successful in analysing data in econometrics (which resulted in 2003 Nobel Prize in Economic Sciences for C. W. J. Granger and R. Engel), finance and engineering \cite{crato,gil-alana,fou}. It is a model which fully describes the behaviour of time series using small number of parameters, which can be estimated from the data using well-established techniques and widely available statistical packages \cite{BDbook,farima}. Moreover, these techniques allow for control of the estimation's quality, checking the correctness of the model or even forecasting future values of the time series. In recent years, new physical \cite{solarFlare, Nose}, biological \cite{burneckiBiol} and medical \cite{MEDarma} applications  of ARFIMA model were found, allowing for empirical description of complex systems with long (power-like), short (exponential) and finite-range dependencies \cite{STarma, farima}, see Fig.  \ref{f:corr}. ARMA processes were also studied as  models of physical data governed by discrete-time Langevin equations \cite{GLEarma, MZarma}.

The main physical interpretation of this model was based on the fact that ARFIMA approximates processes such as fractional Brownian motion, L\'evy stable motions \cite{farima,conv} and the corresponding noises, whereas its special case ARMA can model properties of various stationary processes with finite or exponentially decaying memory \cite{BDbook,beran}. However, most of these continuous-time processes themselves reflect rather behaviour of the process then its internal physical dynamics. On the other hand, the mathematical theory proposed by A. W. Philips \cite{philips}, in recent years developed further by P. Brockwell, R. Davies and Y. Yang \cite{BD}, establish a connection between ARMA model and a class of continuous-time stochastic dynamical systems. Here we show that the physical importance of these results is significant, and after suitable refinement, this connection establishes reliable physical basis for ARMA and ARFIMA models. 
\begin{center}
\textbf{II. ARFIMA MODEL}
\end{center}
The studied model ARFIMA($p,d,q$) states that the considered time series $X_n$ fulfils the recursive relation \cite{farima}
\bgeq\label{eq:ARFIMA}
\Delta^d \lt(X_{n}-\sum_{k=1}^p\phi_kX_{n-k}\rt)= \xi_{n}+\sum_{j=1}^{q}\theta_j \xi_{n-j},
\eeq
where $\phi_k,\theta_j, d$ are deterministic coefficients and $\xi_n$ is white noise which generates the stochastic dynamics; it is Gaussian or non-Gaussian, e.g. $\alpha$-stable \cite{farima, kokoszka} which determines the distribution of $X_n$. The above equation is comprised of three basic building blocks: AR part (left parentheses), FI part (operator $\Delta^d$) and MA part (right side). Each of these blocks models different type of memory and has distinct interpretation. If no memory is present we deal with ARFIMA(0,0,0), which is a white noise: $X_n=\xi_n$.

The left parentheses of Eq. \Ref{eq:ARFIMA} is AR($p$) (Auto Regressive) part  in which coefficients $\phi_k$ determine how the present value of the time series $X_n$ depends linearly on the past values $X_{n-k}$; in other words it models internal dynamics of the system. Because this dynamics is linear, it describes the exponential components of the memory. The most basic process from this class is called AR(1) or ARMA(1,0), which is the simplest exponential memory process with correlation function \cite{BDbook}
\bgeq
\rho_X(k)=\frac{\Braket{X_nX_{n+k}}}{\sqrt{\Braket{X_n^2}\Braket{X_{n+k}^2}}} = \e^{-\phi_1k}.
\eeq
 More general AR($p$) models have memory functions which are mixtures of exponential decays \cite{BDbook}. These processes have great importance for statistics because AR($p$) are maximal entropy processes for fixed first p+1 values of the autocovariance function \cite{entr}.

The right side of Eq. \Ref{eq:ARFIMA} is MA($q$) process (Moving Average) which  determines the external dynamics through modification of the white noise. It models finite-range components of the memory which depend on coefficients $\theta_j$. Actual value of MA($q$) process depends only on $q$ last values of the the generating noise $\xi_n$, and because of that it does not contain any information about the history of the process older then $q\Delta t$. For example, process called MA(1) or ARMA(0,1) is  the simplest type of coloured noise with ultra short memory and correlation function $\rho_{\text{MA(1)}}(1)=\theta_1$, $\rho_{\text{MA(1)}}(k)=0$ for $k>1$ \cite{BDbook, BJbook}.

The last part of Eq. \Ref{eq:ARFIMA}, the operator $\Delta^d$ denoting the FI (Fractional Integration) part reflects both non-stationarity and fractional memory. The symbol $\Delta$ denotes the discrete difference operator: $\Delta X_n=X_{n+1}-X_n$. When $d$ is a natural number, the non-stationary process ARFIMA($p,d,q$), in this case also called ARIMA($p,d,q$), is understood as a process witch after $d$ differentiations is stationary ARMA($p,q$). Basic example is ARFIMA(0,1,0) which is summed white noise, that is sampled trajectory of Brownian motion. 

In a situation when $d$ is real, it can be decomposed into natural-number part $d_n$ and fractional remainder $d_f$, $-1/2\le d_f\le 1/2$, such that $\Delta^{d}=\Delta^{d_n}\Delta^{d_f}$. This remainder accounts for the power-law type of memory common for, e.g., anomalous diffusion \cite{anom}. This operator is understood as a series \cite{farima}
\bgeq
\Delta^{d_f}X_n = \sum_{k=0}^\infty\frac{d_f(d_f-1)\cdots(d_f-k+1)}{k!}(-1)^kX_{n-k}.
\eeq
 Applying $\Delta^{-d_f}$ to both sides of Eq. \Ref{eq:ARFIMA}, it can be confirmed that ARFIMA($p,d,q$) can be interpreted as modification of ARFIMA($p,d_n,q$) generated by noise $\Delta^{-d_f}\xi_n$. This noise, which is denoted as FI($d_f$) or ARFIMA($0,d_f,0$) is a stationary process with power-law memory function witch has a tail $\sim n^{2d_f-1}$; when $\xi_n$ are Gaussian this time series is very similar to discrete-time fractional Brownian noise \cite{conv,barkai, kou}.

\begin{figure}[h!] \centering
\includegraphics[width=8cm]{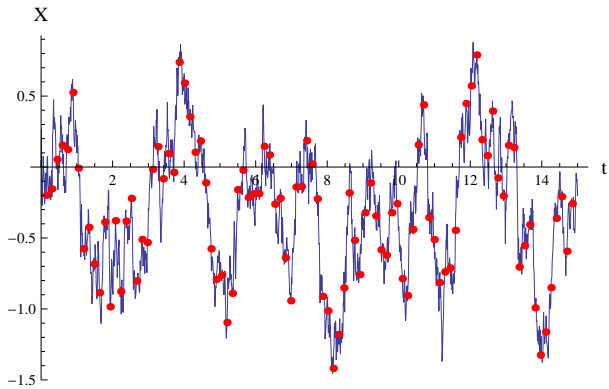}
\caption{(Color online) Continuous-time process (blue line) and sampled process (red dots) for the trajectory of stochastic harmonic oscillator.}
\label{f:sampl}
\end{figure}

\begin{center}
\textbf{III. CONTINOUS- VERSUS DISCRETE-TIME PROCESSES IN EXPERIMENTS}
\end{center}
A~continuous-time process $X(t)$ in a natural way contain a lot more information than its discrete-time counterpart $X_n=X(n\Delta t)$, see Fig. \ref{f:sampl}. During sampling we lose information, e.g. about the geometrical properties of trajectories. Memory functions become discrete and do not contain information about the dependence within intervals smaller than sampling rate $\Delta t$. However, only in discrete time we can define some more refined memory functions, like partial autocorrelation function, which is correlation of $X_n$ and $X_{n+k}$ with the influence of all in-between $X_{n+j}$ removed \cite{BJbook, BDbook}, which in many cases has a simple form which leads to a greater usability. 

Some state functions, like power spectral densities (psd) differ considerably for discrete and continuous time. These are memory functions of the considered process in the Fourier space. Continuous-time psd (cpsd) of the process $X$: $f_X$, is most easily defined as a Fourier transform of the covariance function
\bgeq
f_X(\omega)=\int_{-\infty}^\infty d \tau\ \Braket{X(t)X(t+\tau)}\e^{-\I\omega\tau}.
\eeq
 But, in contrast to psd $f_X$, discrete-time power spectral density (dpsd) is a Fourier series of the sampled covariance function
\bgeq
f^{\Delta t}_{X}(\omega)=\frac{1}{\Delta t}\sum_{k=-\infty}^\infty \Braket{X_{n}X_{n+k}}\e^{-\I\omega \Delta t k},
\eeq
 and is a periodic function witch repeats after $2\pi/\Delta t$. The relation between these two functions is given by Poisson summation formula \cite{spec, spec2}. It allows to calculate dpsd numerically or analytically given cpsd, but for the processes considered in this paper it is not a very practical tool. However, using Poisson summation formula, one can prove that dpsd converges to cpsd as $\Delta t\to 0$ \cite{BJbook, spec}
\bgeq
\lim_{\Delta t\to 0}f_{X}^{\Delta t}=f_X;
\eeq
 this fact can be interpreted as convergence of the time series $X_n$ to the process $X(t)$ as $\Delta t\to 0$ (this limit is often called infill asymptotics). But, in realistic conditions, we often are far away from this limit and only discrete-time model properly reflect the behaviour of the observed system. Note also that all distortions of the data caused by the measurement equipment are changes of the sampled series $X_n$ as this is the object that is actually processed by the hardware. Thus, accounting for unwanted effects like blur or measurement noise must be performed in the discrete time setting \cite{ja}. In order to perform this procedure, the discrete-time model of the undistorted observations is needed.

\begin{figure}[h!] \centering
\includegraphics[width=8cm]{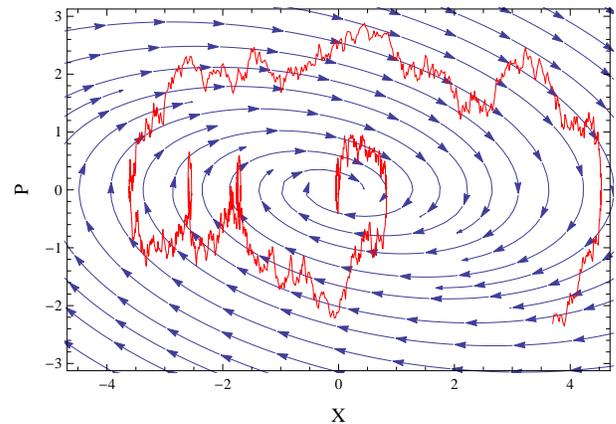}
\caption{(Color online) Phase plot of Eq. \Ref{eq:hp} (blue lines with arrows) for coordinates $X, P$ and stochastic solution of Eq. \Ref{eq:hp} (red line), $m=1,\kappa=1/4,\beta=1/4$.}\label{f:phase}
\end{figure}

\begin{center}
\textbf{IV. LINEAR DYNAMICAL SYSTEMS}
\end{center}
To link these two notions using ARFIMA model we will consider a linear stochastic system with $N$-dimensional state vector $\bdS(t)=[S^1(t),S^2(t),\ldots,S^N(t)]^\mathrm{T}$, which evolves in continuous time according to stochastic differential equation of first order\cite{sobczyk}
\bgeq\label{eq:Sdiff}
\frac{d}{d t} \bdS(t) = A \bdS(t) +\bdF(t),
\eeq
where $A$ is $N\times N$ matrix with constant coefficients and  $\bdF(t)$ is some stationary noise, acting as a random force. This general model describes class of systems with time-independent environment and additive stochastic disturbance. Note,  that if we study state vector $\bdS(t)$ in which some coordinates $S^i(t)$ are described by differential equations of order bigger than one, we can complement the state vector $\bdS(t)$ by auxiliary coordinates $\{\frac{d}{d t} S^i(t), \frac{d^2}{d t^2} S^i(t),\ldots\}$ and also reduce the problem to the form \Ref{eq:Sdiff}. One well-known property described by Eq. \Ref{eq:Sdiff} is a position of particle trapped in the harmonic potential within liquid \cite{lang}
\bgeq\label{eq:hp}
m\frac{d^2}{dt^2}X(t)=-\kappa X(t)-\beta \frac{d}{dt}X(t)+F(t),
\eeq
 where $m$ is the mass of particle, $\kappa$ is stiffness of the harmonic trap, $\beta$ is friction coefficient of the liquid and $F(t)$ is white noise modelling the exchange of momenta with surrounding particles. Phase plot of this equation is shown in Fig. \ref{f:phase}. Other examples include evolution of charge $Q(t)$ in a linear RLC circuit disturbed by noise electromotive force $\mathcal{E}(t)$ 
\bgeq\label{eq:RLC}
L\frac{d^2}{dt^2}Q(t) +R\frac{d}{dt}Q(t) +\frac{1}{C}Q(t)=\mathcal{E}(t),
\eeq
as well as other types of linear disturbed circuits \cite{RLC}, harmonic heat bath models \cite{zwanzig}, Brownian magnetic particle in constant magnetic field \cite{Bmag} and many more.

\begin{center}
\textbf{V. TIME DISCRETISATION PROCEDURE (SAMPLING)}
\end{center}
 In any of these cases the real system evolves in continuous time, but the experimental observations must be discrete and usually have the form of time series $\bdS_n$ sampled with constant sampling time $\Delta t$: $\bdS_n =\bdS(n\Delta t)$. The necessary condition for $\bdS(t)$ and $\bdS_n$ to be stationary is for matrix $A$ to be negative-definite, in other words it needs to have eigenvalues with negative real part. In such case there exists stationary solution of \Ref{eq:Sdiff} given by convolution of the force $\bdF(t)$ with matrix exponential $\e^{At}$ \cite{sobczyk} 
\bgeq\label{eq:Ssol}
\bdS(t)= \int_{-\infty}^t d s\ \e^{A(t-s)}\bdF(s).
\eeq
From elementary properties of the matrix exponential and integration it follows that
\begin{align}
\bdS(n\Delta t)&= \e^{A\Delta t}\int_{-\infty}^{(n-1)\Delta t} \!\!\!\! ds\ \e^{A(t-s)}\bdF(s)\nonumber\\
 &+ \int_{(n-1)\Delta t}^{n\Delta t}\!\!\!\! d s\ \e^{A(n\Delta t-s)}\bdF(s),
\end{align}
 that is the sampled process $\bdS_n$ fulfils the equation
\bgeq\label{eq:Sn}
\bdS_{n}=E\bdS_{n-1}+\bdXi_{n}.
\eeq
The obtained recursive formula is a vector counterpart of the process AR(1) and is called VAR(1) (Vector AR) \cite{BDbook}. The VAR(1) process $\bdS_n$ can be explicitly expressed in terms of the generating noise $\bdXi_n$ as
\bgeq\label{eq:Sdisc}
\bdS_{n}=\sum_{k=0}^\infty E^k\bdXi_{n-k}.
\eeq
The above equation is a discrete counterpart of the convolution formula \Ref{eq:Ssol}.

 From Eq. \Ref{eq:Sdisc} it fallows that neither matrix $A$ directly, nor all values of $\bdF(t)$ affect the state $\bdS_n$, they do this only through their discrete-time counterparts
\bgeq\label{eq:EXi}
E=\e^{A\Delta t},\quad\bdXi_{n}=\int_{(n-1)\Delta t}^{n\Delta t}\!\!\!\! d s\ \e^{A(n\Delta t-s)}\bdF(s).
\eeq
 The left part of the above equation defines the discretized evolution operator $E$, which is responsible for the deterministic part of the transition from state $\bdS_{n-1}$ to $\bdS_n$. The stochastic deviation from the deterministic path is described by the right part of \Ref{eq:EXi}, which defines the packing operator $\bdF(t)\mapsto\bdXi_n$. The whole influence of the process $\bdF(t)$ on $\bdS_n$ is fully determined by the packed force $\bdXi_n$. Each of values $\bdXi_n$ gathers values of $\bdF(t)$ from interval $((n-1)\Delta t,n\Delta t)$. If $\bdF(t)$ is stationary, also the discrete noise $\bdXi_n$ is stationary. The packed force process inherits most of the properties of the underlying continuous-time $\bdF(t)$: the same type of distribution (Gaussian $\bdF(t)$ gives Gaussian $\bdXi_n$, $\alpha$-stable $\bdF(t)$ gives $\alpha$-stable $\bdXi_n$) and the same type of memory (white noise, finite range, exponential or power law). 

 Because of the above correspondence, the statistical methods available for VAR(1) model \cite{pfaff} can be used in physical applications. Using estimators for matrix $E$ and taking matrix logarithm we obtain estimates for the underlying matrix $A=\ln(E)/\Delta t$. The properties of force $\bdF(t)$ can be studied through analysis of $\bdXi_n$, which can be estimated as $\bdXi_n=\bdS_{n}-E\bdS_{n-1}$. 

\begin{center}
\textbf{VI. SINGLE COMPONENT ANALYSIS}
\end{center}
The above approach has a huge disadvantage: it requires that we can observe the whole sampled state vector $\bdS_n$, which is unrealistic for more complex systems. To avoid this difficulty we want to analyse the behaviour of one chosen component of $\bdS_n$, without loss of generality we will assume it is $S_n^1$. The evolution of this component is coupled with the evolution of the rest of state vector by action of non-diagonal components of the evolution matrix~$E$.

In order to decouple component $S_n^1$, we use Cayley-Hamilton theorem, which states that if $N\times N$ matrix $E$ has characteristic equation $p(\lambda) =\det(E-\lambda I)= \lambda^N-\sum_{k=1}^N \phi_k \lambda^{N-k} = 0$, then the matrix $E$ itself fulfils this equation, that is
\bgeq\label{eq:pM}
p(E)  = E^N-\sum_{k=1}^N \phi_k E^{N-k}=0.
\eeq
This polynomial equation of order $N$ has coefficients $\phi_k$ determined by the eigenvalues $\lambda_i$ of the matrix $E$
\bgeq
\phi_k=(-1)^{k+1}\sum_{D_k}\prod_{i\in D_k}\lambda_i,
\eeq
 where $D_k$ denotes family of all $k$-element subsets of the set $\{1,2,\ldots, N\}$. The eigenvalues $\nu_i$ of matrix $A$ from \Ref{eq:Sdiff} and the eigenvalues of matrix $E$ are related by formula $\lambda_i=\e^{\nu_i\Delta t}$. Therefore $E$ fulfils \Ref{eq:pM} with coefficients 
\bgeq
\phi_k=(-1)^{k+1}\sum_{D_k}\e^{\Delta t\sum_{i\in D_k}\nu_i}.
\eeq

As a preparation for using Cayley-Hamilton theorem, we recursively use \Ref{eq:Sn} and express variables $\bdS_{n-k}$ as functions of $\bdS_{n-N}$
\begin{align}
\bdS_{n-N\phantom{+1}}&=\phantom{E^2}\bdS_{n-N}\nonumber\\
\bdS_{n-N+1}&=E\phantom{^2}\bdS_{n-N}+\phantom{E^2}\bdXi_{n-N+1}\nonumber\\
\bdS_{n-N+2}&=E^2\bdS_{n-N}+E\phantom{^2}\bdXi_{n-N+1}+\phantom{E}\bdXi_{n-N+2}\nonumber\\
\bdS_{n-N+3}&=E^3\bdS_{n-N}+E^2\bdXi_{n-N+1}+E\bdXi_{n-N+2}+\bdXi_{n-N+3}\nonumber\\
\phantom{\bdS_{n-N+2}}&\ \ \vdots\nonumber\\
\bdS_{n\phantom{-N+1}} &= E^{N}\bdS_{n-N}+\sum_{j=0}^{N-1}E^j\bdXi_{n-j},
\end{align}
thus coupling them to only this one variable. After multiplying these equations by $\phi_k$ in order to obtain terms $\phi_kE^{N-k}\bdS_{n-N}$, and subtracting them from the last one, we remove all action of the matrix $E$ on the time series $\bdS_n$ using equality $(E^N-\sum_{k=1}^N \phi_k E^{N-k})\bdS_{n-N}=0$ obtained from the Cayley-Hamilton theorem. The cost of this decoupling is performing complicated transformations of the discretized force $\bdXi_n$ along the way. The equation that we obtain after this procedure has a form
\bgeq\label{eq:Snv}
\bdS_n-\sum_{k=1}^N\phi_k \bdS_{n-k} = \bdXi_{n}+\sum_{k=1}^{N-1}R_{k}\bdXi_{n-k}.
\eeq
The left side of the above equation is AR($N$) part described by coefficients $\phi_k$, which depend only on the deterministic matrix $A$. The left side acts as an effective noise $\boldsymbol{\eta}_n$
\bgeq
\boldsymbol{\eta}_n=\bdXi_{n}+\sum_{k=1}^{N-1}R_{k}\bdXi_{n-k}
\eeq
which generates the stochastic dynamics of the vector $\bdS_n$. The behaviour of this noise is determined by the matrices $R_k$
\bgeq
R_k = E^k-\sum_{j=1}^k\phi_jE^{k-j}
\eeq
composed of mixtures of time-shifting operators $E^{k-j}$. They affect the evolution by mixing different $\bdXi_{n-k}$; as a result the first component $S_n^1$ fulfils the recurrence relation
\bgeq
S_n^1-\sum_{k=1}^N\phi_k S_{n-k}^1 =\eta_n^1.
\eeq

There is a deep connection between the above formula and the classical Mori-Zwanzig theory \cite{zwanzig,mori}. The above equation can be written in slightly different manner
\bgeq
\frac{\Delta S_n^1}{\Delta t}-\sum_{k=1}^N\phi_k' S_{n-k}^1 =\frac{\eta_n^1}{\Delta t},
\eeq
where $\phi_1'=(\phi_1+1)/\Delta t, \phi_k'=\phi_k/\Delta t,k>1$. Now it becomes clear that this is the discrete-time analogue of the generalised Langevin equation  \cite{kou, zwanzig,mori}. It is no coincidence, as derivations in both cases use the same ideas, decoupling most of the coordinates of freedom, at the same time introducing the memory kernel and effective noise. Despite many similarities, this analogy has its limitations, e.g. there is no clear discrete-time equivalent of the fluctuation-dissipation theorem. 
\begin{center}
\textbf{VII. ANALYSIS OF THE STOCHASTIC FORCE}
\end{center}
The effective noise $\eta_n^1$ is a mixture of $N-1$ last values and all components of $\bdXi_n$. If the underlying force $\bdF(t)$ was white Gaussian noise, then the resulting $\eta_n^1$ are a mixture of Gaussian white noises which forms a time series with $N-1$ finite-range dependence. When analysing only one component, we can ignore its internal structure and represent it as MA($N-1$) process
\bgeq\label{eq:eta}
\eta_n^1 = \xi^1_{n}+\sum_{j=1}^{N-1}\theta^1_j \xi^1_{n-j},
\eeq
generated by a white noise $\xi^1_n$, which is in fact orthogonalised series $\eta^1_n$ \cite{MAortog}. Such orthogonalisation can be always performed and coefficients $\theta_j^1$ may be obtained solving system of equations resulting from comparing the covariance function of left side of Eq. \Ref{eq:eta} and effective noise $\eta_n^1$ \cite{BDbook}.

 Thus, we have come to conclusion that coefficient $S_n^1$ is ARMA($N,N-1$) process with coefficients $\phi_k$ and $\theta_j^1$. Similar statements hold for more complex models of the force $\bdF(t)$.

If the force $\bdF(t)$ has finite range of memory smaller than $K\Delta t$, then $\eta_n^1$ by the same reasoning as above can be regarded as MA($N-1+K$) process. In such case $S_n^1$ is ARMA($N,N-1+K$).

If the force $\bdF(t)$ has power-law memory tails $\sim t^{2d_f-1}$, then $\eta_n^1$ is a composition of finite-range mixing introduced by operators $R_k$ and the power-law behaviour. The best approximation of such time series is FIMA($d_f,q$), i.e. process similar to \Ref{eq:eta}, but where $\xi_n^1$ are FI($d_f$) time series \cite{farima}. So, component $S_n^1$ is ARFIMA($N,d_f,q$).

If the force $\bdF(t)$ has exponential tails of memory, that is, if it can be represented in form similar to \Ref{eq:Ssol}, then we may treat it as a time-dependent state of the same class as $\bdS(t)$, which confirms that $\eta_n^1$ is ARMA($L,L-1+N$) for some $L$; $+N$ results from mixing of $\bdXi_{n-k}$. The AR($L$) part, understood as an operator, can be freely moved from acting on $\eta_n^1$ to acting on $S_n^1$, leading to ARMA($N+L,N+L-1$) model. The operator AR($N+L$) can be easily calculated as composition AR($L$)AR($N$).

In our considerations we assumed that $\bdS(t)$ was stationary. But, we may observe the possibly non-stationary integral of the stationary coordinate $X(t)=\int_0^td \tau\ S^1(\tau)$, $X(t)$ being position, charge, etc. In this case  the process of differences $\Delta X_n = \int_{(n-1)\Delta t}^{n\Delta t}d \tau\ S^1(\tau)$ is ARFIMA($p,d_f,q+1$) which follows from the fact that $S^1(\tau+n\Delta t)$ is some ARFIMA($p,d_f,q$) for any $\tau$, with $p,d_f,q$ determined by the proper model from the description above. Increase by one order in MA part accounts for additional short-time memory introduced by the integral $\int_{(n-1)\Delta t}^{n\Delta t}d s$. And so, $X_n$ is ARFIMA($p,d_f+1,q+1$). 

Analogical statement holds for all other components $S^i_n$. The AR coefficients $\phi_k$ are identical for all of them,  MA coefficients $\theta^i_j$ and noises $\xi^i_n$ from Eq. \Ref{eq:eta} vary. The equations governing evolution of different $S^i_n$ were decoupled, however these components are dependent, because $\xi_n^i$ for different $i$ are mixtures of $\bdXi_n$ components, so they are dependent set of variables.

We stress that all modelling is performed at the level of the discretized stochastic force, the obtained ARFIMA model of the observed coordinate is by no means phenomenological as is often the case of discrete-time models, but derived from the theory of the continuous-time dynamical system. For all cases except the power-law memory the correspondence is exact, in the latter case FIMA approximation must be made for the discretized stochastic force, AR part is still exact. As FI part can reflect any type of power-law long-time memory asymptotics and MA part can account for any finite-range deviations, such model is most often sufficient \cite{burneckiBiol,farima}.

\begin{figure}[h!] \centering
\includegraphics[width=8cm]{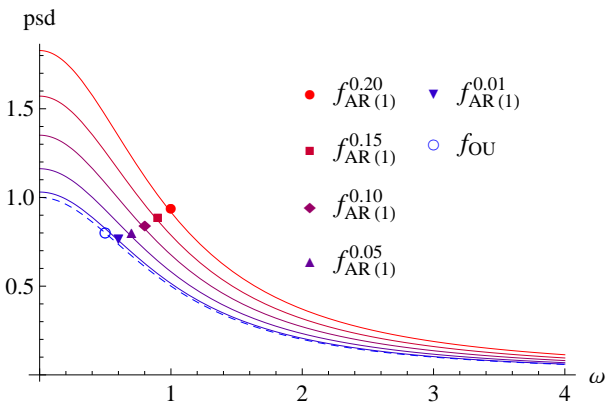}
\caption{(Color online) Comparison of dpsd and cpsd of the process \Ref{eq:OU} for decreasing sampling times $\Delta t$ in non-dimensional units, $\kappa=\beta=\sigma=1$.}\label{f:psd}
\end{figure}

In our reasoning we used the fact that Gaussian process is fully determined by its covariance structure at the moment when we orthogonalised the effective noise series $\eta_n^1$. For non-Gaussian processes it is no longer true, as they can have richer than linear memory structure \cite{kokoszka}. Therefore, for non-Gaussian forces $\bdF(t)$, the obtained ARFIMA model reflect only linear aspect of the memory. In this case it is approximate, but has the same autocovariance and power spectral density as the original process.
\begin{center}
\textbf{VIII. PARTICLE IN A HARMONIC POTENTIAL} 
\end{center}
Let us come back to Eq. \Ref{eq:hp}, which is the second order differential equation describing the particle trapped in harmonic potential. An approximation, in which the inertial term $m\, d^2X/dt^2$ is considered negligible, simplifies analysis. In such conditions the state of particle evolves according to force-balance equation $\beta\, d X/dt =-\kappa X+\xi$. Its stationary solution,
\bgeq\label{eq:OU}
X(t)=\frac{1}{\beta}\int_{-\infty}^t d s\ \e^{-\frac{\kappa}{\beta}(t-s)}\xi(s)
\eeq
is called Ornstein-Uhlenbeck process \cite{OU,protter}, and has a well-known Lorenzian continuous-time power spectral density \cite{lang,ja}
\bgeq\label{eq:cpsd}
f_{\text{OU}}(\omega) = \frac{1}{\beta^2}\frac{\sigma^2}{\big(\frac{\kappa}{\beta}\big)^2+\omega^2},
\eeq
 where $\sigma^2$ is the variance of the noise $F(t)$; when fluctuation-dissipation theorem holds $\sigma^2=2k_BT\beta$ \cite{zwanzig}. Sampled trajectory of \Ref{eq:OU} is AR(1) process with a coefficient $\phi_1=\e^{-\Delta t \kappa/\beta}$. The discrete-time power spectral density can be calculated from the general formula for all ARFIMA processes \cite{spec, beran}, which for AR(1) yields \cite{ja}
\bgeq
f^{\Delta t}_{\text{AR(1)}}(\omega)=\left(\phi_1^{-2}-1\right)\frac{1}{2 \kappa\beta}\frac{\sigma^2\Delta t}{1+\phi_1^2-2\phi_1\cos(\omega\Delta t)}.
\eeq
As we see cpsd and dpsd functions differ when $\Delta t$ is not considerably smaller then $\beta/\kappa$ (see Fig. \ref{f:psd}), which is often the case for mesoscopic objects observed in normal conditions \cite{ja}. As we provide exact formula of the observed spectral density, there is no reason to use approximate Eq. \Ref{eq:cpsd}.

In a case when the mass is not negligible, the effects of sampling become more complex. The full state vector $\bdS$ is then composed of position and momentum, $\bdS(t) = [X(t), P(t)]^\mathrm{T}$. The stochastic force affects the change of momenta $\bdF(t) = [0, F(t)]^\mathrm{T}$. Note that because of the identical form of Eq. \Ref{eq:hp} and Eq. \Ref{eq:RLC}, all further results would follow also for the RLC circuit \Ref{eq:RLC} after a simple change of letters. In this case the state vector would consists of charge and electric current. 

Calculating the eigenvalues
\bgeq
\nu_{1,2}=-\frac{\beta}{2m}\pm\sqrt{\left(\frac{\beta}{2m}\right)^2-\frac{\kappa}{m}}
\eeq
 of the evolution matrix $A = \bigl[\begin{smallmatrix} 0&1/m\\ -\kappa&-\beta/m \end{smallmatrix}\bigr]$, we obtain the AR coefficients of the sampled position process $X_n$ which, if $F(t)$ is a white noise, is ARMA(2,1) with AR(2) coefficients
\begin{align}
\phi_1&=2\exp\left(-\Delta t\frac{\beta}{2m}\right)\cosh\lt(\Delta t\sqrt{\left(\frac{\beta}{2m}\right)^2-\frac{\kappa}{m}}\rt)\nonumber,\\
\phi_2&=-\exp\left(-\Delta t\frac{\beta}{m}\right).
\end{align}
The MA coefficient $\theta^1_1$ is also determined by the calculated eigenvalues and is given by complicated formula, but can be easily calculated numerically. Estimating the AR(2) coefficients from the data, the ratios $\kappa/m$ and $\beta/m$ of the underlying process can be assessed and the parameter $m$ can be subsequently estimated from variance of the sampled process $X_n$.  If $F(t)$ is not white noise, the MA part may differ and if $F(t)$ would have the power-law dependence it would be reflected in the FI part of ARFIMA model. 

\begin{center}
\textbf{IX. SUMMARY} 
\end{center}

In our work we tried to construct a bridge between continuous-time linear dynamical systems and discrete-time ARMA or, more generally, ARFIMA models. The studied correspondence for many cases might serve as physical interpretation of the ARFIMA model and justification for its usage. Additionally, we have shown what order physical ARFIMA model should have for given dynamical system and we have given explicit formulas for its AR coefficients, which allows for estimation the dynamical system's parameters using standard statistical tools. Its MA and FI coefficients can also be calculated, but they depend on the assumed model of the stochastic force. The coefficients of ARFIMA model determine its characteristics, such as power spectral density, linking the basic dynamical system model with functions that can be estimated from the sampled data measured during experiment.

\begin{acknowledgements}
\begin{center}
\textbf{ACKNOWLEDGEMENTS} 
\end{center}
This research was supported by NCN Maestro Grant No. 2012/06/A/ST1/00258.
\end{acknowledgements}

\end{document}